\documentstyle[12pt]{article}
\def\lappeq{\mathrel{\rlap{\raise.5ex\hbox{$<$}}
{\lower.5ex\hbox{$\sim$}}}}
\begin{document}
\begin{titlepage}
\pagestyle{empty}
\baselineskip=21pt
\rightline{Alberta Thy-38-92}
\rightline{CfPA 92-035}
\rightline{UMN-TH-1114/92}
\rightline{November 1992}
\vskip .2in
\begin{center}
{\large{\bf Inflation, Neutrino Baryogenesis, and (S)Neutrino-Induced
Baryogenesis}}
\end{center}
\vskip .1in
\begin{center}
Bruce A. Campbell

{\it Department of Physics, University of Alberta}

{\it  Edmonton, Alberta, Canada T6G 2J1}

Sacha Davidson

{\it Center for Particle Astrophysics, University of California}

{\it Berkeley, California, 94720 USA}

Keith A. Olive

{\it Department of Physics and Astronomy, University of Minnesota}

{\it Minneapolis, MN 55455, USA}

\vskip .1in

\end{center}
\centerline{ {\bf Abstract} }
\baselineskip=18pt

\noindent
%%%%%%%%%%%%%%%%%%%%%%%%%%%%%%%%%%%%%%%%%%%%%
%%%%%%%%%%%%%%%%%%%%%%%%

%%%%%%%%%%
We evaluate the constraints
that the COBE observations put on baryogenesis in inflationary
cosmologies.We consider the supersymmetric version of the
 proposal of Fukugita and Yanagida, that
the
baryon asymmetry of the universe is created by nonperturbative
electroweak reprocessing of a lepton asymmetry generated in the decay
of heavy right handed see-saw (s)neutrinos.
We review our recent proposal of a
mechanism for baryogenesis via sphaleron reprocessing of a lepton
asymmetry generated by (s)neutrino mass effects on flat
direction scalar condensate oscillations. Finally we analyze in
detail the implementation of these mechanisms in the recently
proposed
ansatze for fermion mass matrices in supersymmetric, unified,
theories.
%%%%%%%%%%%%%%%%%%%%%%%%%%%%%%%%%%%%%%%%%%%%%
%%%%%%%%%%%%%%%%%%%%%%%%
 \vskip .1in

\centerline{ {Accepted For Publication In: {\it Nucl.Phys. B}} }

%%%%%%%%%%
\end{titlepage}
\baselineskip=18pt
{\newcommand{\la}{\mbox{\raisebox{-.6ex}{$\stackrel{<}{\sim}$}}}
{\newcommand{\ga}{\mbox{\raisebox{-.6ex}{$\stackrel{>}{\sim}$}}}
%\newpage
%%%%%%%%%%%%%%%%%%%%%%%%%%%%%%%%%%%%%%%%%%%%%
%%%%%%%%%%%%%%%%%%%%%%%%

%%%%%%%%%%
%%%%%%%%%%%%%%%%%%%%%%%%%%%%%%%%%%%%%%%%%%%%%
%%%%%%%%%%%%%%%%%%%%%%%%

%%%%%%%%%%
\section{Introduction}

     Baryogenesis \cite{sak,kt} is a crucial element
 in cosmological and particle physics
models, and plays a key role in constraining
their form.
First, the model must be capable of generating a baryon asymmetry.
Second, it must be capable of preserving the asymmetry despite the
possible dangers of subsequent entropy generation at cosmological
phase transitions, or the equilibration of baryon number violating
interactions. Models of
inflation \cite{infl} have always been constrained by baryogenesis.
``Sufficient"(for baryogenesis and nucleosynthesis) reheating after
inflation is a requirement for any successful inflationary scenario.
Non-perturbative standard model electroweak interactions \cite{KRS1}
also
constrain the structure of possible unification models, because of
the danger that a previously established baryon asymmetry may be
washed out \cite{KRS1,am}. In addition, the potential erasure of the
baryon
asymmetry by sphaleron effects leads to constraints on new
interactions violating baryon or lepton number
\cite{fy2,ht,bn,cdeo,fglp,iq,dr}.

	The fact that baryon number violating interactions
 in the standard model are relatively unsuppressed
\cite{KRS1} at high
 temperatures,
 has focussed a great deal of attention on attempts
 to generate the cosmological baryon asymmetry entirely
 within the standard model at the relatively low temperature of
0(100 GeV)
\cite{S,Mc,D}. Recently, more detailed calculations of the
effective potential at finite temperatures \cite{D,C} have shown however,
 that the baryon asymmetry is not generated unless the Higgs mass
 is \(m_H~\la~ 50 GeV\),  in contradiction with
experimental
 limits \cite{lep}.  Even in the minimal supersymmetric extension
 of the standard model, the possibility looks doubtful \cite{G},
 though non-minimal extensions with a more complicated Higgs
 sector \cite{Piet} may turn out to provide viable models of
baryogenesis.

       Among the simple mechanisms for primordial (before the
electroweak scale) baryogenesis are the out of equilibrium decay of a
heavy gauge/higgs boson \cite{wei,ttwz}, and the decay of a sfermion
condensate
oscillating along a flat direction in a supersymmetric theory
as proposed by Affleck and Dine \cite{ad}.
Another possibility, proposed by Fukugita and Yanagida
\cite{fy1}(hereafter  FY), is the out of
equilibrium decay of a heavy neutrino (see also \cite{lut}). In this
case sphaleron effects
are required to transform the lepton asymmetry produced in the decay
into a baryon asymmetry. In addition the heavy Majorana neutrino
whose decay is responsible for baryogenesis in this mechanism, could
induce, via its exchange, a lepton number violating dimension-five
effective interaction. If the reactions induced by this operator were
in thermal equilibrium after the decay of the $N_R$, then they would
combine with sphaleron interactions to erase the lepton and baryon
asymmetry as Fukugita and Yanagida realized \cite{fy2}.
Demanding that the erasure not occur then
constrains the induced interaction, and hence the neutrino masses.
This double-edged behaviour, that any interactions violating baryon
and lepton number, which could in principle cause
baryo(lepto)-genesis if out of equilibrium, could equally well erase
the BAU if they equilibrate, allows us to derive limits on these
interactions from the persistence of a primordial baryon (lepton)
asymmetry.

 Recently we have proposed an alternate mechanism by which
(s)neutrino mass terms in a supersymmetric theory can generated a BAU
\cite{cdo1}(hereafter CDO). This mechanism also depends on the
non-perturbative electroweak reprocessing of a lepton asymmetry.
However, in this case, the lepton asymmetry to be reprocessed is
generated by the effects of lepton number violating operators,
induced by the see-saw (s)neutrino masses, acting on scalar
condensate oscillations along flat directions (before supersymmetry
breaking) of the supersymmetric theory.

 In this paper we wish to explore the connection between
inflation and these mechanisms of baryogenesis. In particular we note
that the recent COBE observations \cite{cobe} of the anisotropy of
the cosmic
microwave background radiation can be related to the reheat
temperature after inflation in generic models. We show that inflation
may be capable of softening the previously derived constraints on
baryon and lepton number violating interactions, inferred from the
persistence of a primordial BAU. We then discuss the implications of
the derived reheat temperature on models of baryogenesis.We discuss
the FY mechanism, and calculate the BAU generated in its extension to
supersymmetric models. We discuss the mechanism proposed in CDO,
which is a new
mechanism for baryogenesis from neutrino Majorana masses, arising in
supersymmetric theories. Finally we consider the implementation of
both
the FY mechanism and the mechanism proposed in CDO via the recently
proposed
ansatze for
(s)neutrino mass matrices, and discuss the relation to neutrino
physics and
astrophysics.
%%%%%%%%%%%%%%%%%%%%%%%%%%%%%%%%%%%%%%%%%%%%%
%%%%%%%%%%%%%%%%%%%%%%%%

%%%%%%%%%%%%%%%%%%%%%%%%%%%%%%%%%%%%%%%%%%%%%
%%%%%%%%%%%%%%%%%%%%%%%%

%%%%%%%%%%%%%%%%%%%%%%

\section{Constraints From Inflation}

       Consider a generic scalar potential of the form:
%%%%%%%%%%%%%%%%%%%%%%%%%%%%%%%%%%%%%%%%%%%%%
%%%%%%%%%%%%%%%%%%%%%%%%
%%%%%%%%%%
\begin{equation}
 V( \eta ) = {{\mu}^4} P( \eta )
\end{equation}
%%%%%%%%%%%%%%%%%%%%%%%%%%%%%%%%%%%%%%%%%%%%%
%%%%%%%%%%%%%%%%%%%%%%%%
%%%%%%%%%%
where $\eta$ is the scalar field driving inflation, the inflaton;
$\mu$ is an as yet unspecified mass parameter, and $P(\eta)$ is a
function of $\eta$ which possesses the features necessary for
inflation, but contains no small parameter. That is, we expect
possible cubic or quartic interactions to be of the form
${\lambda_3}\mu^4{(\eta/{M_P})^3}$ or
${\lambda_4}\mu^4{(\eta/{M_P})^4}$, with
$\lambda_{3,4}\sim 0(1)$ and $M_P$ the Planck mass. Examples of
potentials of this form are: ``Mexican hat" potentials \cite{ab}
$V={\mu^4}(1-{\eta^2/{M_P^2}})^2$ where
${\lambda_4}=1$; potentials used for chaotic
inflation \cite{chao} $V=({\lambda_4}/4)(\mu/M_P)^4{\eta^4}$ or
$V=(1/2){m^2}{\eta^2}$ where
${m^2}={2{\mu^4}}/{{M_P}^2}$; Pseudo-Goldstone Boson (PGB) potentials
\cite{pseudo}
$V={{\mu^4}{\cos}{\theta}/{f_{\theta}}}$, which for
${f_\theta}\sim{M_P}$ can be expanded about $\theta=0$, so that
$V\simeq{\mu^4}(1-{{\theta^2}/{M_P^2}}+\ldots)$ exactly as in the
expansion of the ``Mexican hat" potential above; potentials derived
from minimal $N=1$ supergravity superpotentials \cite{nost}
$F={\mu^2}(1-{\eta}/{M_P})^2M_P$; or potentials derived from no-scale
supergravity superpotentials \cite{eenos}
$F={\mu^2}({\eta}-{\eta^4}/{4M_P^3})$. Clearly most of the useful
inflationary potentials can be put into the form of Equation (1).

       For large scale fluctuations of the type measured by COBE
\cite{cobe}, ie.
for those which ``reenter" the horizon in the matter dominated epoch,
density perturbations at horizon crossing are given by \cite{hwk}:
       %%%%%%%%%%%%%%%%%%%%%%%%%%%%%%%%%%%%%%%%%%%
%%%%%%%%%%%%%%%%%%%
%%%%%%%%%%%%%%%%%
\begin{equation}
{\frac{\delta\rho}{\rho}\simeq\frac{H^2}{10{\pi^{3/2}}\dot{\eta}}
\simeq
 O(100)\frac{\mu^2}{M_P^2}}
\end{equation}
%%%%%%%%%%%%%%%%%%%%%%%%%%%%%%%%%%%%%%%%%%%%%
%%%%%%%%%%%%%%%%%%%%%%%%
%%%%%%%%%%
where the exact coefficient obviously depends on the specific
potential under consideration (for chaotic inflation from a quadratic
potential, the coefficient is somewhat smaller). The magnitude of the
density fluctuations can be related to the observed quadropole
\cite{pee}
moment:
%%%%%%%%%%%%%%%%%%%%%%%%%%%%%%%%%%%%%%%%%%%%%
%%%%%%%%%%%%%%%%%%%%%%%%
%%%%%%%%%%
\begin{equation}
{\langle{a_2^2}\rangle = \frac{5}{6}
2{\pi^2}{(\frac{\delta\rho}{\rho})^2}}
\end{equation}
%%%%%%%%%%%%%%%%%%%%%%%%%%%%%%%%%%%%%%%%%%%%%
%%%%%%%%%%%%%%%%%%%%%%%%
%%%%%%%%%%
(we are here assuming that scalar perturbations are dominant
\cite{dhsst}). The
observed quadropole moment gives \cite{cobe}:
%%%%%%%%%%%%%%%%%%%%%%%%%%%%%%%%%%%%%%%%%%%%%
%%%%%%%%%%%%%%%%%%%%%%%%
%%%%%%%%%%
\begin{equation}
{\langle{a_2^2}\rangle = (4.7\pm2)\times{10^{-10}}}
\end{equation}
%%%%%%%%%%%%%%%%%%%%%%%%%%%%%%%%%%%%%%%%%%%%%
%%%%%%%%%%%%%%%%%%%%%%%%
%%%%%%%%%%
or
%%%%%%%%%%%%%%%%%%%%%%%%%%%%%%%%%%%%%%%%%%%%%
%%%%%%%%%%%%%%%%%%%%%%%%
%%%%%%%%%%
\begin{equation}
{\frac{\delta\rho}{\rho} = (5.4\pm1.6)\times{10^{-6}}}
\end{equation}
%%%%%%%%%%%%%%%%%%%%%%%%%%%%%%%%%%%%%%%%%%%%%
%%%%%%%%%%%%%%%%%%%%%%%%
%%%%%%%%%%
which in turn fixes the coefficient $\mu$ of the inflaton potential:
%%%%%%%%%%%%%%%%%%%%%%%%%%%%%%%%%%%%%%%%%%%%%
%%%%%%%%%%%%%%%%%%%%%%%%
%%%%%%%%%%
\begin{equation}
{\frac{\mu^2}{M_P^2} = few\times{10^{-8}}}
\end{equation}
%%%%%%%%%%%%%%%%%%%%%%%%%%%%%%%%%%%%%%%%%%%%%
%%%%%%%%%%%%%%%%%%%%%%%%
%%%%%%%%%%

       Fixing $({\mu^2}/{M_P^2})$ has immediate general consequences
for inflation \cite{eeno}. For example, the Hubble parameter during
inflation,
${{H^2} \simeq (8\pi/3)({\mu^4}/{M_P^2})}$ so that $H \sim
10^{-7}M_P$. The duration of inflation is $\tau \simeq
{M_P^3}/{\mu^4}$, and the number of e-foldings of expansion is $H\tau
\sim 8\pi({M_P^2}/{\mu^2}) \sim 10^{9}$. If the inflaton decay rate
goes as $\Gamma \sim {m_{\eta}^3}/{M_P^2} \sim {\mu^6}/{M_P^5}$, the
universe recovers at a temperature $T_R \sim (\Gamma{M_P})^{1/2} \sim
{\mu^3}/{M_P^2} \sim 10^{-11} {M_P} \sim 10^8 GeV$. Recall that
before COBE all that could be set was an upper limit on $\mu$.

       The low reheat temperature and low inflaton mass ${m_\eta}
\simeq {\mu^2}/{M_P} \simeq 10^{11} GeV$ can be quite constraining on
models of baryogenesis. In the out of equilibrium decay scenarios,
this means that the Higgs bosons in question must have masses $M_H
\leq m_{\eta} \sim 10^{11} GeV$.
To have Higgs triplets with baryon and lepton number violating
interactions this light is problematic for proton stability. Although
the Yukawa couplings of these triplet Higgs are related by group
theory factors to the Yukawas of the doublet Higgs responsible for
the quark and lepton masses, and hence small for the first generation
particles appearing in proton decay diagrams, nonetheless, masses for
the triplet Higgs of this order would, in many models, cause proton
decay at experimentally disallowed rates. The problem goes from
troubling to terminal in simple susy-GUTs, where the relatively light
Higgs triplets are accompanied by their Higgsino superpartners.
Exchange of these particles then induces dimension-five effective
interactions, suppressed by only one power of the Higgsino mass. When
dressed by external gaugino exchange, one gets a proton decay
amplitude down by only one power each of the Higgsino mass, and the
chargino (or neutralino or gluino) mass. For gaugino masses of order
a TeV or less, this induces proton decay at a disallowed rate unless
the Higgsino mass is of order the susy-GUT unification scale, and
hence useless for baryogenesis after inflation. It is possible to
avoid the dimension five operators, at the price of complicating the
model. For example, in the SU(5) susy-GUT the dimension five operator
is induced by the diagram of Figure 1, where the cross denotes the
mixing of the Higgs 5 and $\bar{5}$ representations. If one were to
duplicate the Higgs representation content, so that there were two
Higgs 5 and two Higgs $\bar{5}$ representations, where one of each
coupled to give the standard masses to the quarks and leptons, and
the duplicate ``sterile" representation of each did not couple to
quarks and leptons, and if one arranged that the mass mixing between
the 5 and $\bar{5}$ representations was such that the normal 5 only
mixed with the ``sterile" $\bar{5}$ and vice versa, then one would
have effectively killed the dimension five operator, but at the price
of an ad hoc extension of the Higgs sector. Another issue that would
need to be addressed in any susy-GUT featuring Higgs triplets below
$10^{11}$ GeV is the effect of these states and their superpartners
on the renormalization group equations governing the running of the
gauge couplings. While undesired effects on the beta functions for
the gauge running might in principle be compensated for by the
effects of the addition of other multiplets, the procedure again
lacks outside motivation.

       Another possibility for generating the baryon asymmetry is the
decay of sfermion condensates in a susy-GUT \cite{ad}. Here the
baryon number
is generated by the oscillation of scalar fields along a flat
direction of the scalar potential \cite{ad,lin85}. When the dilution
of the baryon to
entropy ratio due to inflaton decay has been taken into account, the
baryon asymmetry is given by \cite{eeno}:
%%%%%%%%%%%%%%%%%%%%%%%%%%%%%%%%%%%%%%%%%%%%%
%%%%%%%%%%%%%%%%%%%%%%%%
%%%%%%%%%%
\begin{equation}
{\frac{n_B}{s} \simeq
\frac{\delta{\phi_o^4}{m_{\eta}^{3/2}}}{{M_X^2}{M_P^{5/2}}{\tilde{m
}}}}
\end{equation}
%%%%%%%%%%%%%%%%%%%%%%%%%%%%%%%%%%%%%%%%%%%%%
%%%%%%%%%%%%%%%%%%%%%%%%
%%%%%%%%%%
where $\delta$ is a combination of coupling constants
parametrizing the CP-violation in the sfermion decay, $\phi_o$
is the initial sfermion vev, which is determined by quantum
fluctuations during inflation, so that ${{\phi_o}^2} \simeq
{H^3\tau}/{4{\pi}^2} \simeq {\mu}^2$ and $\tilde{m} \simeq 10^{-16}
M_P$ is the supersymmetry breaking scale, and $M_X \simeq 10^{-3}
M_P$ is the unification scale. Then the baryon asymmetry becomes:
%%%%%%%%%%%%%%%%%%%%%%%%%%%%%%%%%%%%%%%%%%%%%
%%%%%%%%%%%%%%%%%%%%%%%%
%%%%%%%%%%
\begin{equation}
{\frac{n_B}{s} \simeq
\frac{\delta{\mu^7}}{{M_X^2}{M_P^{4}}{\tilde{m}}} \simeq 10^{-4} \delta}
\end{equation}
%%%%%%%%%%%%%%%%%%%%%%%%%%%%%%%%%%%%%%%%%%%%%
%%%%%%%%%%%%%%%%%%%%%%%%
%%%%%%%%%%

       Of course, with regard to sphalerons, the Affleck-Dine
mechanism based on supersymmetric SU(5) fares no better than the out
of equilibrium decay scenario based on SU(5). The operators involving
flat direction fields (LQQQ) with non-vanishing vevs preserve B-L and
therefore will have any asymmetry they produce above $T_c$ (the
electroweak phase transition temperature) erased. Though it may be
possible to push down the temperature after the condensate decay, the
typical reheat temperature is of order $10^4 GeV > T_c$. This may not
be
terribly problematic in unified groups beyond SU(5), as they may
induce baryon number violating operators which don't conserve B-L,
and involving flat direction fields \cite{ceno}\cite{morg}. What this
does suggest, however, is that in the
simpler models, we should look for baryogenesis mechanisms that are
immune to sphaleron erasure. For these reasons, in the next two
sections we turn to baryogenesis mechanisms induced by neutrino
Majorana mass terms. We first turn to the FY mechanism, involving the
out of equilibrium decay of heavy Majorana right handed neutrinos,
and derive its supersymmetric extension. In the succeeding section we
turn to the mechanism proposed in CDO, which is only possible in
supersymmetric
theories, and in which lepton number violating potential terms,
induced in the low energy potential (after supersymmetry breaking)
from the (s)neutrino Majorana mass superpotential terms, act on
scalar condensate oscillations along flat directions of the potential
(before supersymmetry breaking). In section five we then examine the
implementation of these mechanisms in recently proposed ansatze for
neutrino masses in supersymmetric theories, to assess their
effectiveness. The advantage of these two mechanisms is that since
they are driven by B-L violating Majorana neutrino masses, the
non-zero B-L produced is reprocessed to baryons, and sphaleron
effects actually cause them to function, rather than posing a
danger to the BAU they produce.

%%%%%%%%%%%%%%%%%%%%%%
 \section{Neutrino Baryogenesis}

         In this section we would like to discuss
the
possibility of generating the baryon asymmetry by the decay of a
massive right handed see-saw (s)neutrino, in a supersymmetric model.
In fact, the possibility of generating the
baryon
asymmetry by the decay of superheavy leptons predates \cite{dvn} the
suggestion
of Fukugita and Yanagida \cite{fy1}.  In these
 earlier works however, the heavy
lepton was presumed to decay via baryon number violating GUT
interactions, into two or three body final states with non-zero
baryon number. For leptons light enough to be produced by
thermalization or inflaton decay after inflation, it is difficult to
arrange baryon number violating decays without the relatively light
fields responsible destabilizing the proton. What is novel and
elegant in the Fukugita-Yanagida proposal, is that the heavy lepton
decay directly produces only a lepton asymmetry, and may be arranged
to do so with physics as simple as a see-saw neutrino mass matrix.
The baryon asymmetry is then subsequently produced by sphaleron
reprocessing of the lepton asymmetry, since $L\neq0$ from the heavy
neutrino decay implies $B-L\neq0$, and the equilibrium condition then
results in a BAU.

If one adds heavy gauge singlet majorana neutrinos to the Standard
Model via
 the Lagrangian (in two-component notation (uppercase letters for fermions)
and in four component notation (lower case))
%%%%%%%%%%%%%%%%%%%%%%%%%%%%%%%%%%%%%%%%%%%%%
%%%%%%%%%%%
\begin{eqnarray}
 i {N}^{\dagger k}  \partial_{\mu} \sigma^{\mu}  N^k
- {N}^k  M_k N^k
 + \lambda_{kj}  H {N}^k  L^j +
h.c. \nonumber \\
= \frac{1}{2} {\bar n}^k (i\not\partial - M_k) n^k +
\lambda_{kj}  H {\bar n}^k P_L l^j + h.c.
 \label{S1}
\end{eqnarray}
%%%%%%%%%%%%%%%%%%%%%%%%%%%%%%%%%%%%%%%%%%%%%
%%%%%%%%%%%%%
where the majorana masses $M_k$ are large, then below the electroweak
phase
transition the SU(2) doublet neutrinos acquire majorana masses of
order
$\lambda^2 <\! H \!>^2/M$ (seesaw). Such light neutrinos could solve
the
solar neutrino problem or consitute hot dark matter, and,  as
suggested by
Fukugita and Yanagida, the heavy $N^i$ could produce a lepton
asymmetry when
 they decay in the
early Universe.

To produce an LAU,  one needs out-of-equilibrium $L$, $C$ and $CP$
violating interactions \cite{sak}. The majorana mass (or the higgs
coupling,
depending
on whether one assigns lepton number to the $N^i$) violates $L$, the
Universe expansion prevents the decay from being in exact thermal
equilibrium,
 and the phases in the matrix $\lambda$ (combined with the imaginary
part of
the tree-loop amplitude) provide $CP$ violation. If one defines, as a
measure
of the $CP$ violation in a decay
\begin{equation}
\epsilon = \frac{\Gamma - \Gamma_{CP}}{\Gamma + \Gamma_{CP}}
\label{S2}
\end{equation}
then the lepton asymmetry produced in the out-of-equilibrium decays
is $\Delta L \sim 10^{-2} \epsilon$.  For
 the decay
 $N^i \rightarrow H L^j$, in which case the CP violation comes from
the
interference between the tree diagram and figure 2a,  this
has been calculated to be \cite{fy1,lut}
%%%%%%%%%%%%%%%%%%%%%%%%%%%%%%%%%%%%%%%%%%%%%
%%%%%%%%%%%%%%
\begin{equation}
\epsilon_i =  \frac{1}{2 \pi (\lambda
\lambda^{\dagger})_{ii}}
 \sum_j \left(Im[(\lambda \lambda^{\dagger})_{ij}]^2 \right)
f(M_j^2/M_i^2)  \label{S3}
\end{equation}
%%%%%%%%%%%%%%%%%%%%%%%%%%%%%%%%%%%%%%%%%%%%%
%%%%%%%%%%%%%%%
where
\begin{equation}
f(x) = \sqrt{x} ( 1 - (1 + x) \ln \left[ \frac{1+x}{x} \right] )~~. \label{ssf}
\end{equation}

The supersymmetric generalization of (\ref{S1}) would be to add
kinetic terms
for the gauge singlet  superfields (denoted boldface) ${\bf N}^i$,
and
superpotential terms
\begin{equation}
M_i {\bf N}^i {\bf N}^i + \lambda_{ij}{\bf N}^i {\bf H} {\bf L}^j +
h.c.
\label{S5}
\end{equation}
to the Lagrangian for the minimal supersymmetric standard model.
We have chosen a basis for the ${\bf N}^i$ such that the mass
matrix
is diagonal
and real. The ${\bf N}^i$ in this notation are left-handed, so would
be
the CP-conjugates of the heavy right-handed neutrinos of the
see-saw mass
matrix. In component field notation, where we use tildes over
scalars (Higgs excepted) to distinguish them from their partner
fermions,
(\ref{S5}) gives quartic scalar interactions that will not contribute
to
$CP$ violation at one loop, and
\begin{eqnarray}
- \lambda_{ij} M_i \tilde{N}^{i*} H \tilde{L}^j + \lambda_{ij}
\tilde{N}^i
\tilde{H} L^j
+
\lambda_{ij} H {N}^i  L^j + \lambda_{ij}\tilde{L}^j
{\tilde{H}} N^i
+h.c. \nonumber \\
= - \lambda_{ij} M_i \tilde{N}^{i*} H \tilde{L}^j + \lambda_{ij}
\tilde{N}^i
{\bar h}^c P_L l^j
+
\lambda_{ij} H {\bar n}^i P_L l^j + \lambda_{ij}\tilde{L}^j
{\bar h}^c P_L n^i
+h.c.
\label{S6}
\end{eqnarray}
 The heavy
neutrinos ($N^i$)
  therefore can decay to $L^j H$ or to $\tilde{L}^j
{\tilde{H}}$, and the
sneutrinos to
$ \tilde{L}^j H$ or $L^j {\tilde{H}}$. The tree level decay
rate for each
of
these processes
is
\begin{equation}
\Gamma_D^i = \frac{( \lambda \lambda^{\dagger})_{ii} M_i}{16 \pi}~~.
\label{S7}
\end{equation}
The one loop diagrams which will contribute to the CP violating
parameter $\epsilon$
in each of these decays are listed in figure 2. Since this is
a supersymmetric
theory, one might worry that these diagrams cancel; fortunately
however, they are contributions to the
D-term ${\bf N}^*{\bf HL}$, so superpotential non-renormalization
theorems do
not
apply
and a non-zero lepton asymmetry is possible. The CP-violating
parameters
for the four decays are equal,
so the lepton asymmetry due to the out-of-equilibrium decays of the
$i$th
generation
neutrino and sneutrino is (assuming a maximally out-of-equilibrium
decay)
%%%%%%%%%%%%%%%%%%%%%%%%%%%%%%%%%%%%%%%%%%%%%
%%%%%%%%%%%%%%%%%%%%%%%
\begin{equation}
\frac{n_L}{n_\gamma} \simeq  - 10^{-2} \frac{1}{2 \pi
(\lambda \lambda^{\dagger})_{ii}} \sum_j \left( Im [
(\lambda \lambda^{\dagger})_{ij}]^2 \right) g(M_j^2/M_i^2)
 \label{S8}
\end{equation}
where
\begin{equation}
g(x) = 4 \sqrt{x} \ln \frac{1 + x}{x} \label{ssg}
\end{equation}
%%%%%%%%%%%%%%%%%%%%%%%%%%%%%%%%%%%%%%%%%%%%%
%%%%%%%%%%%%%%%%%%%%%%%%
After sphaleron interactions are included this turns into a baryon
asymmetry,
%%%%%%%%%%%%%%%%%%%%%%%%%%%%%%%%%%%%%
\begin{equation}
\frac{n_B}{n_\gamma} \simeq -\frac{28}{79} \frac{n_L}{n_\gamma}
\end{equation}
%%%%%%%%%%%%%%%%%%%%%%%%%%%%%%%%
where $n_L/n_\gamma$ here is from (\ref{S8}).

As FY themselves  noted \cite{fy2}, integrating the heavy neutrinos
(and
sneutrinos) out of the Lagrangian (\ref{S1}) ((\ref{S6}) in the
supersymmetric
case)  leads to an
effective L violating interaction of the form
%%%%%%%%%%%%%%%%%%%%%%%%%%%%%%%%
\begin{equation}
\frac{\lambda^2}{M}LLHH  \label{S9}
\end{equation}
%%%%%%%%%%%%%%%%%%%%%%%%%%%%%%%%
which must be out of equilibrium for the baryon (lepton) asymmetry to
survive. Thus we must require the rate
%%%%%%%%%%%%%%%%%%%%
\begin{equation}
\Gamma_I \simeq \frac{\zeta(3)}{8\pi^3} \frac{\lambda^4 T^3}{M^2}
\end{equation}
%%%%%%%%%%%%%%%%%%%%%%%%%%%%%%%%
associated with (\ref{S9}) to be out of equilibrium leading to the
constraint \cite{fy2,ht,cdeo,fglp}
%%%%%%%%%%%%%%%%%%%%%%%%%%%%%%%
\begin{equation}
\frac{M}{\lambda^2} ~\ga~ 1.4 \times 10^{-2} \sqrt{T_{B-L} M_P}
\label{x}
\end{equation}
%%%%%%%%%%%%%%%%%%%%%%%%%%%%%%%%
The temperature scale in (\ref{x}) is to be understood as the lowest
temperature
at which the B-L asymmetry is produced, or $T_m \sim 10^{12}$  GeV,
the maximum
temperature for which sphaleron interactions are in equilibrium.
In the FY scenario above, $T_{B-L}$ is just the reheat temperature
subsequent to N decay, $T_N$,
%%%%%%%%%%%%%%%%%%%%%%%%%%%%%%%%%%%%%%
\begin{equation}
T_{B-L} = T_N \simeq \sqrt{\frac{\Gamma_DM_P}{25}}
\end{equation}
%%%%%%%%%%%%%%%%%%%%%%%%%%%%%%%%
so that
%%%%%%%%%%%%%%%%%%%%%%%%%%%%%%
\begin{equation}
\frac{M}{\lambda^2} ~\ga~ min[3 \times 10^{-4} \lambda^{4/3}M_P,
1.4 \times 10^{-2}(T_mM_P)^{1/2}]
\label{y}
\end{equation}
%%%%%%%%%%%%%%%%%%%%%%%%%%%%%%%%

The above analyses was carried out without considering the effects of
inflation.
In an inflationary model, one must require first that the
right-handed
neutrino can be produced by inflaton decays, $M < m_\eta \sim 10^{11}
GeV$.
The lower bound on M in (\ref{y}) is also modified. Now,
%%%%%%%%%%%%%%%%%%%%%%%%
\begin{equation}
\frac{M}{\lambda^2} ~\ga~ min[3 \times 10^{-4} \lambda^{4/3}M_P,
1.4 \times 10^{-2}(\tilde T_R M_P)^{1/2}]
\label{z}
\end{equation}
%%%%%%%%%%%%%%%%%%%%%%%%%%%%%%%%
where $\tilde T_R$ is related to the inflationary ``reheat"
temperature $T_R \sim \mu^3/M_P^2 \sim 10^8 GeV$;
the limit (\ref{z}) requires equilibrium to be established, and
thermalization after inflation occurs only at a temperature
$\tilde T_R \sim \alpha^2 T_R$ \cite{eeno} where $\alpha$
is a typical gauge coupling constant.  Thus $\tilde T_R \sim 10^5 GeV$.
Then the lepton asymmetry after N decay
(provided $\lambda$ is not too small) is
%%%%%%%%%%%%%%%%%%%%%%%%%%%%%%%%%%%%%%
\begin{equation}
\frac{n_L}{n_\gamma} \simeq \frac{n_\eta}{n_\gamma}\epsilon \sim
(\frac{m_\eta}
{M_P})^{1/2}\epsilon \sim \frac{\mu}{M_P}\epsilon
\end{equation}
%%%%%%%%%%%%%%%%%%%%%%%%%%%%%%%%%
What is especially important about the above modification due to
inflation
is that it applies to all of the constraints on dimension five or
greater
B and L violating operators combined with sphaleron interactions.
The constraints on renormalizable operators are all determined by the
equilibrium condition at $T = T_c \sim O(100) GeV$.  Furthermore,
the softening of the constraint due to inflation does not depend on
supersymmetry.
This is in contrast to the important realization by Ibanez and
Quevedo \cite{iq} of a
softening in supersymmetric models due to the presence of additional
anomalies.  The supersymmetric softening is less severe than
the one described here, ie. $T_{max} \sim 10^8 GeV$, due to
supersymmetric anomalies.
%%%%%%%%%%%%%%%%%%%%%%%%%%%%%%%%%%%%%%%%%%%%%
%%%%%%%%%%%%%%%%%%%%%%%%
\section{(S)Neutrino-Induced Baryogenesis}

       In this section we review the mechanism of baryogenesis
recently
proposed in CDO \cite{cdo1},
arising from the presence of see-saw neutrino masses in
supersymmetric theories. We first
derive the contributions to the
low energy effective potential of a supersymmetric theory, induced
after
supersymmetry breaking, by the
singlet (s)neutrino interactions. We then examine the
effect of these interactions on slepton and squark condensates
oscillating along ``flat directions" (before supersymmetry breaking)
of the low-energy supersymmetric standard model. We show that these
interactions act on the condensate oscillations to produce a net
lepton asymmetry, which nonperturbative electroweak effects partially
reprocess
into baryons. The mechanism resembles that of
Fukugita and Yanagida in that it
requires a see-saw neutrino mass, and in that the dynamical
mechanism only generates a lepton asymmetry at first, with sphalerons
responsible for partially reprocessing that into baryons.

              On the other hand, unlike the proposal of Fukugita and
Yanagida, the dynamics of this mechanism involves the oscillations of
sfermion condensates along susy flat directions, and hence
unlike theirs can only occur in supersymmetric extensions of the
standard model. Furthermore, in this mechanism the CP violation
necessary for the production of the asymmetry may arise spontaneously
from the phase of the condensate, and hence is naturally of order
unity, whereas in the Fukugita-Yanagida scenario one needs hard CP
violation in the neutrino-Higgs Yukawa couplings. Finally, as this
mechanism depends on the effective low-energy interactions induced by
the right handed see-saw neutrinos, it can be operative even when the
singlet neutrino masses are too large for them to be physically produced in the
post-inflationary epoch, whereas in the Fukugita-Yanagida mechanism
they must be copiously produced, thus bounding their mass by the
inflaton mass scale, which COBE results give as $\leq 10^{11} GeV$
for typical inflationary models, as shown above.

       In order to have this mechanism of baryogenesis one needs
the following elements. First we need to show the existence of flat
(before supersymmetry breaking) directions in the potential of our
model, including the contributions coming from the extra
superpotemtial terms involving the singlet neutrino. The motion, after
supersymmetry breaking potentials turn on, of the scalar condensates
(squark, slepton, and Higgs) along these directions  drives the lepton
asymmetry generation. Second, we need to establish the existence of
slepton number violating potential interactions, induced after
supersymmetry breaking by the neutrino mass see-saw superpotential
terms, which pick up a non-zero contribution along the flat
direction, and which act during the course of the scalar oscillations
to build up a net slepton density. Third, we must follow the
evolution of the condensate to calculate the lepton asymmetry
produced, and its subsequent dilution by inflaton decay, to get the
final lepton asymmetry for sphaleron reprocessing.

       To demonstrate the mechanism, we need to exhibit flat
directions in the supersymmetric standard model, extended to include
neutrino see-saw masses which arise from the superpotential
(\ref{S5}), and renormalizable $N$-field superpotential terms
$k_1^{i} {\bf N}^i {\bf H}_1{\bf H}_2 + k_2^{ijk}{\bf N}^i {\bf N}^j {\bf N}^k
$. Note that the
singlet neutrino  interactions  in this section therefore differ from those in
the
previous one; the superpotential terms involving ${\bf N^i}$ are
\begin{equation}
M_i {\bf N}^i {\bf N}^i + \lambda_{ij}{\bf N}^i {\bf H} {\bf L}^j +
k_1^{i} {\bf N}^i {\bf H}_1{\bf H}_2 + k_2^{ijk}{\bf N}^i {\bf N}^j {\bf N}^k +
h.c.
\end{equation}
  The
potential from these and the usual standard model  F-terms (plus the
SU(3)xSU(2)xU(1) D-terms) has, among others, the following flat
direction
(which is a generation permuted version of one appearing in reference
\cite{morg}); it depends
on three arbitrary complex parameters $a,v,c,$ and four phases
$\alpha,\beta,
\phi$, and $\gamma$. We work in a generation basis in which the
${g_e^{ij}}$ and the ${g_d^{ij}}$ have been diagonalized; the quark
indices denote quark colour.
%%%%%%%%%%%%%%%%%%%%%%%%%%%%%%%%%%%%%%%%%%%%%
%%%%%%%%%%%%%%%%%%%%%%%%
%%%%%%%%
\begin{equation}
\begin{array}{ccc}
{{\tilde{t}}^c_3}=a & {{\tilde{t}}^1}=v &
{{\tilde{\nu}}_e}={e^{i\gamma}}c\\
{{\tilde{b}}^c_3}=c &{{\tilde{s}}^c_2}={e^{i\alpha}}
\sqrt{{|a|}^2+{|c|}^2}& \\
{{\tilde{\mu}}^-}={e^{i\beta}} \sqrt{{|v|}^2+{|c|}^2}
&{{\tilde{d}}^c_1}={e^{i\phi}} \sqrt{{|a|}^2+{|v|}^2+{|c|}^2}& \\
\end{array}
\end{equation}
%%%%%%%%%%%%%%%%%%%%%%%%%%%%%%%%%%%%%%%%%%%%%
%%%%%%%%%%%%%%%%%%%%%%%%
%%%%%%%%
A vev along this particular flat direction produces a
non-zero vev for the effective scalar operator
%%%%%%%%%%%%%%%%%%%%%%%%%%%%%%%%%%%%%%%%%%%%%
%%%%%%%%%%%%%%%%%%%%%%%%
%%%%%%%%
\begin{equation}
\langle{{\tilde{\mu}}^-}{{\tilde{\nu}}_e}{{\tilde{b}}^c_3}{({{\tilde{
t}}^c_3})^*}\rangle = {e^{i(\beta+\gamma)}}{a^*}{c^2}
\sqrt{{|v|}^2+{|c|}^2}
\label{a}
\end{equation}
%%%%%%%%%%%%%%%%%%%%%%%%%%%%%%%%%%%%%%%%%%%%%
%%%%%%%%%%%%%%%%%%%%%%%%
%%%%%%%%
which violates lepton number by two units.

       After supersymmetry breaking, this scalar operator will be
induced by the singlet (s)neutrino interactions via the diagram of Figure
3. In the diagram the insertions on the $\tilde{N}$ line and vertex  are the
supersymmetry breaking scalar mass and interaction A-terms $(O(m_{\delta}$)).
The
resulting potential term coupling is of order
$V=\lambda{{\phi}{\phi}{\phi}{{\phi}^*}}$,
where ${{\phi}{\phi}{\phi}{{\phi}^*}}$ corresponds to the quartic
scalar operator of equation (\ref{a}) and
%%%%%%%%%%%%%%%%%%%%%%%%%%%%%%%%%%%%%%%%%%%%%
%%%%%%%%%%%%%%%%%
%%%%%%%%%%%%%%%
\begin{equation}
\lambda \simeq \frac{{g_{\nu}^e}{g_{\nu}^{\mu}}{g_b}{g_t}}{(4 \pi)^3}
\frac{m_{\delta}^2M^2}{(M^2 +g^2 \phi_o^2)^2}
\label{b}
\end{equation}
%%%%%%%%%%%%%%%%%%%%%%%%%%%%%%%%%%%%%%%%%%%%%
%%%%%%%%%%%%%%%%%%%%%%%%
%%%%%%%%
where ${g_{\nu}^e}$ and ${g_{\nu}^{\mu}}$ are the (experimentally
undetermined) neutrino
see-saw Dirac mass Yukawas, and M is the scale of the large singlet
${\bf N }$ Majorana mass term (We assume $k_1, k_2 \sim 1)$. This estimate for
the scale of the
induced quartic scalar coupling is supported by the general arguments
of
\cite{morg} for operators of this form.

The calculation of the lepton (and baryon) asymmetry
produced by the sfermion condensate was undertaken in
CDO \cite{cdo1}.  If we
denote the expectation values, after inflation, of scalars
parametrizing the flat directions as $\phi_o
=\langle{o|\phi|o}\rangle$, producing $V_o = \langle{o|V|o}\rangle$,
then we can then write the net lepton number per scalar particle
associated
with the oscillations of $\phi$ as
%%%%%%%%%%%%%%%%%%%%%%%%%%%
\begin{equation}
L \sim \frac{Im V_o}{{m_\delta}^2 {\phi_o}^2} \sim \frac{\theta \lambda
{\phi_o}^4}
{{m_\delta}^2 {\phi_o}^2} \sim
 O(10^{-5}) \theta g_{\nu}^e g_{\nu}^\mu \frac{{\phi_o}^2 M^2}{(M^2 + g^2
\phi_o^2)^2}
\label{c}
\end{equation}
where $\theta \sim 1$ is the degree of CP violation in (\ref{a})
and we have assumed that $g_t \sim 1$ and
$g_b \sim O(10^{-2})$.
The net lepton number density ($\sim$ the net baryon density after
sphaleron
reprocessing) is then given by
%%%%%%%%%%%%%%%%%%%%%%%%%%%%
\begin{equation}
n_B \sim n_L \sim L m_\delta {\phi_o}^2 {(R_\phi/R)}^3
\label{f}
\end{equation}
%%%%%%%%%%%%%%%%%%%%%%%%%%%%
where R is the cosmological scale factor and $R_\phi$ is the value
of the scale factor when the sfermion oscillations begin.

To evaluate the BAU produced after inflation by this mechanism, we
must recall that the initial value of $\phi_o$ is determined by
quantum fluctuations during
inflation, and that the  the asymmetry is diluted by inflaton decays.
 The initial sfermion expectation value is
 ${\phi_o}^2 \simeq H^3\tau/4\pi$ where the Hubble parameter  $H
\simeq
\mu^2/M_P$ and
the duration of inflation is $\tau \simeq {M_P}^3/\mu^4$ so that
$\phi_o \simeq \mu$, where our discussion above of COBE indicates
$\mu^2 \simeq
few \times 10^{-8} M_P^2$.
The final baryon asymmetry can then be found from \cite{eeno} to be
%%%%%%%%%%%%%%%%%%%%%
\begin{equation}
\frac{n_B}{n_\gamma} \simeq O(10^{-5}) \frac{\theta g_{\nu}^e
g_{\nu}^\mu {\phi_o}^4 M^2
 {m_\eta}^{3/2}}{(M^2 +g^2 \phi_o^2)^2 {M_P}^{5/2} m_\delta} \simeq  O(10^{-2})
\frac{
\theta g_{\nu}^e g_{\nu}^\mu \mu^3 M^2}{{M_P}^4m_\delta}
\label{g}
\end{equation}
%%%%%%%%%%%%%%%%%%%%%%%%%%%%%%%%%%%%%%%%%
where $m_\eta \simeq \mu^2/M_P$ is the inflaton mass, and we have assumed $ M
\la g\phi_o $ and $g^4 \sim 10^{-3}$ in the denominator .  This gives a
 value of $\simeq 10^{-10}$ for $\theta g_{\nu}^e g_{\nu}^\mu
{M}^2/M_P^2 \simeq 10^{-13}$, which  we will evaluate below for realistic
models.

In summary, this mechanism for baryogenesis in supersymmetric
extensions of the standard model, does not involve (super)GUT
interactions. It depends on non-perturbative electroweak reprocessing
of a lepton asymmetry, which is in turn generated by the
effects of lepton number violating induced operators, acting on
scalar condensate oscillations along flat directions of the standard
model. In our realization of the mechanism we have shown that lepton
number violating operators of this type can be induced, after
supersymmetry breaking, by  singlet neutrino interactions that include see-saw
neutrino masses. In principle, any other superpotential interaction inducing
violation of
either
baryon or lepton number has the potential to induce
baryogenesis via sfermion condensate dynamics, coupled with sphaleron
reprocessing, in a manner similar to that of the example we have
presented. In the next section we consider the concrete
implementation of this scenario utilizing neutrino Majorana masses as
discussed above, in the context of recently proposed ansatze for
(s)neutrino mass matrices in supersymmetric theories. We compare the
BAU produced via this mechanism to that from the Fukugita-Yanagida
mechanism, in the same class of theories.
%%%%%%%%%%%%%%%%%%%%%%%%%%%%%%%%%%%%%%%%%%%%%
%%%%%%%%%%%%%%%%%%%%%%%%
%%%%%%%%%%%%%%%%%%%%%%
\section{(S)Neutrino Masses}

       In this section we wish to address the question of whether the
mechanisms discussed in the last two sections provide sufficient
baryogenesis in the context of realistic models of neutrino masses.
By realistic models, we understand that the resulting framework
should be theoretically well motivated, predictive, and giving
experimentally succesful predictions for fermion masses. In the
modern context, this usually means finding a ground state solution to
heterotic superstring theory, that is well motivated and
phenomenologically viable. This is, of course, an exceptionally
ambitious undertaking, as such a solution would encompass all of
particle physics, and despite vigorous efforts at expanding the space
of known string constructions, at present no solution is known that
is completely viable phenomenologically. In view of this, we defer
the task of such a construction, and adopt the strategy of working
with predictive ansatze for the fermion masses that have been
incorporated in grand unified models \cite{hrr}, and that when
renormalization group extrapolated from unification scales to
accessible scales in supersymmetric theories give experimentally
interesting predictions \cite{dhr}. In general we shall be agnostic
with regard to the origin of our fermion mass matrices. As they were
first proposed in the context of an SO(10) grand unified theory \cite
{hrr}, with its accompaning GUT gauge and (rather extensive) Higgs
sectors, we will consider the physical implications of such an origin
for them. However, we will also consider how the physics would appear
if we only retain the pattern of the mass matrices as a
phenomenological ansatze, and introduce no gauge or Higgs
representations or interactions beyond those of the supersymmetric
standard model. We expect that these two, rather extreme, approaches
will bracket the range of models one might reasonably expect. In
particular, string constructions might be expected to lie somewhere
between the extremes, as in general both the gauge and Higgs sectors
are less populous in string models than GUTs, (but more populous
than the susy standard model) due to the possibility of gauge
symmetry breaking by ``twisting" in string models.

       The specific pattern or ``texture" of fermion masses that we
will assume for our considerations is that proposed by Harvey, Ramond
and Reiss in an SO(10) GUT \cite{hrr}(hereafter HRR). Their fermion
mass texture includes an up quark mass matrix incorporating the
Fritzsch ansatze \cite{frit}, and down and charged-lepton mass
matrices incorporating the Georgi-Jarlskog ansatze \cite{geojarl};
the resulting pattern of quark masses yields Oakes-type relations for
the mixing angles. As it was implemented in an SO(10) model, where
neutrinos of both chiralities are incorporated in the 16
representations of fermions, it also predicts a see-saw pattern for
the neutrino masses \cite{seesaw}. In particular the $\Delta I_W=1/2$
Dirac mass terms for the neutrinos are related to the up quark mass
matrices, and $\Delta I_W=0$ Majorana masses for right handed
neutrinos are constrained by the SO(10) Higgs representation content
and the discrete symmetries used to enforce the texture. In order to
be able to enforce the fermion mass texture by discrete symmetries
HRR utilized a rather extensive Higgs sector representation content
(we here give the formulation of the model that directly extends to
the supersymmetric version, and which is equivalent to the form in
their original paper), including a complex 10 and three separate
$\bar{126}$ in the (superpotential) Yukawas, as well as a 54 for GUT
symmetry breaking. The superpotential Yukawas responsible for masses
for the standard model fermion 16s are:
       %%%%%%%%%%%%%%%%%%%%%%%%%%%%%%%%%%%%%%%%%%%
%%%%%%%%%%%%%%%%%%%
%%%%%%%%%%%%%%%
\begin{equation}
L_Y=({A}{16_1}{\cdot}{16_2}+{B}{16_3}{\cdot}{16_3}){\cdot}\bar{126_1}
   +({a}{16_1}{\cdot}{16_2}+{b}{16_3}{\cdot}{16_3}){\cdot}{10}
   +({c}{16_2}{\cdot}{16_2}){\cdot}\bar{126_2}
   +({d}{16_2}{\cdot}{16_3}){\cdot}\bar{126_3}
   \label{superp}
\end{equation}
%%%%%%%%%%%%%%%%%%%%%%%%%%%%%%%%%%%%%%%%%%%%%
%%%%%%%%%%%%%%%%%%%%%%%%
%%%%%%%%
The discrete symmetry responsible for the HRR texture of fermion
masses is spontaneously broken at the unification scale by the Higgs
vevs, and hence as one extrapolates to lower energies by the
renormalization group equations, the renormalization group mixing
will not preserve the texture, modifying the form of the mass
matrices at lower energies. Recently, in a series of interesting
papers, Dimopoulos, Hall and Raby \cite{dhr}(DHR)have shown that if
one uses the renormalization group equations appropriate for the
supersymmetric extension of the standard model, then starting with a
unification scale texture of the HRR form one generates at the
electroweak scale a pattern of quark and lepton masses that is in
agreement with present experiment, and which gives testable
predictions for future experiments. They have also analyzed the
extrapolation of neutrino masses for a specific case of the HRR
neutrino mass texture, when only one vev contributes to a given entry
of the neutrino mass matrix, and have analyzed the resulting
predictions for solar and terrestrial neutrino physics. We will find
interesting features of the DHR restriction on neutrino masses of the
HRR form, when we discuss the FY mechanism.

        Let us now examine neutrino mass matrices of the HRR form. As
there are two neutral Weyl spinors per family of 16, there are a
total of 6 neutral lepton spinors. Of these three are the standard
$SU(2)_W$ doublet neutrinos of the standard model, and three are
singlets. The structure of the 6x6 neutral fermion mass matrix is
thus:
       %%%%%%%%%%%%%%%%%%%%%%%%%%%%%%%%%%%%%%%%%%%
%%%%%%%%%%%%%%%%%%%
%%%%%%%%%%%%%%%
\begin{equation}
M = \left(
    \begin{array}{cc}
     {{M}^{(1)}}&{{M}^{(1/2)}}\\
     {{{M}^{(1/2)}}^T}&{{M}^{(0)}}
    \end{array}
    \right)
\end{equation}
%%%%%%%%%%%%%%%%%%%%%%%%%%%%%%%%%%%%%%%%%%%%%
%%%%%%%%%%%%%%%%%%%%%%%%
%%%%%%%%
where the $M^{\Delta{I_W}}$ are 3x3 generation matrices, whose
superscript represents their $\Delta{I_W}$ value. Following HRR we
assume that there are no $\Delta{I_W}=1$ Higgs expectation values, so
that $M^{(1)}=0$. The isosinglet mass $M^{(0)}$ is given by the vev
of the $\bar{{126}_1}$ along the $\Delta{I_W}=0$ direction.
      %%%%%%%%%%%%%%%%%%%%%%%%%%%%%%%%%%%%%%%%%%%
%%%%%%%%%%%%%%%%%%%
%%%%%%%%%%%%%%%
\begin{equation}
{M^{(0)}} = \left(
            \begin{array}{ccc}
            {0}&{\bar{A}}&{0}\\
            {\bar{A}}&{0}&{0}\\
	    {0}&{0}&{\bar{B}}
            \end{array}
            \right)
\label{k1}
\end{equation}
%%%%%%%%%%%%%%%%%%%%%%%%%%%%%%%%%%%%%%%%%%%%%
%%%%%%%%%%%%%%%%%%%%%%%%
%%%%%%%%
where $\bar{A}=Ake^{i\zeta}$, and $\bar{B}=Bke^{i\zeta}$, with
$ke^{i\zeta}$
the $\Delta{I_W}=0$
vev of $\bar{126_1}$.

Also following HRR, we take (with some minor amendments)
%%%%%%%%%%%%%%%%%%%%%%%%%%%%%%%%%%%%%%%%%%%%%
%%%%%%%%%%%%%%%%%%%%%%%%
\begin{equation}
{M^{(1/2)}} = \left(
            \begin{array}{ccc}
            {0}&\bar{U}&{0}\\
            \bar{U}&{0}&-3\bar{Q}\\
	    {0}&-3\bar{Q}&\bar{W}
            \end{array}
            \right)
\label{k2}
\end{equation}
%%%%%%%%%%%%%%%%%%%%%%%%%%%%%%%%%%%%%%%%%%%%%
%%%%%%%%%%%%%%%%%%%%%%%%
where
$\bar{U}=ape^{i\delta}-3Ate^{i\sigma}$,$\bar{W}=bpe^{i\delta}-3Bte^{i
\sigma}$,
and $\bar{Q}=dqe^{i\mu}$ where $a, A, b, B$, and $d$, are couplings from
the
superpotential (\ref{superp}), and $te^{i\sigma}$ is the
$\Delta{I_W}=(1/2)$
vev of the $\bar{{126}_1}$, $qe^{i\mu}$ is the $\Delta{I_W}=(1/2)$
vev of the
$\bar{{126}_3}$, and $pe^{i\delta}$ is the $\Delta{I_W}=(1/2)$ vev of
the
complex $10$ (appearing in the Higgs doublet with the same weak
hypercharge).
Since in the low-energy effective supersymmetric theory, we expect
these mass
terms to arise from the Yukawa coupling to a single Higgs doublet of
the
required weak hypercharge (which evidently must be a linear
combination of the
complex $10$, $\bar{{126}_1}$, and $\bar{{126}_3}$, in the ratios of
$p, t$, and
$q$), we may write the mass matrix in terms of the vev of that Higgs in
the form:
%%%%%%%%%%%%%%%%%%%%%%%%%%%%%%%%%%%%%%%%%%%%%
%%%%%%%%%%%%%%%%%%%%%%%%
\begin{equation}
{M^{(1/2)}} = \left(
            \begin{array}{ccc}
            {0}&-3u&{0}\\
            -3u&{0}&-3x\\
	    {0}&-3x&-3w
            \end{array}
            \right)
 \frac {v}{\sqrt{2}} \sin \beta
\label{k2p}
\end{equation}
%%%%%%%%%%%%%%%%%%%%%%%%%%%%%%%%%%%%%%%%%%%%%
%%%%%%%%%%%%%%%%%%%%%%%%
where $u$, $x$, and $w$ are chosen so that the mass matrix
(\ref{k2p}) agrees
with (\ref{k2}).

  The heavy
right-handed neutrino mass matrix is made real and diagonal by a
redefinition of
the right handed neutrino fields: $N=UN'$, where
%%%%%%%%%%%%%%%%%%%%%%%%%%%%%%%%%%%%%%%%%%%%%
%%%%%%%%%%%%%%%%%
\begin{equation}
U= {1 \over \sqrt{2}}
       \left(
        \begin{array}{ccc}
         ie^{i\theta}&e^{i\theta}&0  \\
        -ie^{i\theta}&e^{i\theta}&0    \\
          0&0&\sqrt{2}e^{i\phi}
        \end{array}
      \right)
\label{k4}
\end{equation}
%%%%%%%%%%%%%%%%%%%%%%%%%%%%%%%%%%%%%%%%%%%%%
%%%%%%%%%%%%%%%%%%
with $\bar{A}=|Ak|e^{-2i\theta}$, and $\bar{B}=|Bk|e^{-2i\phi}$,
yielding
%%%%%%%%%%%%%%%%%%%%%%%%%%%%%%%%%%%%%%%%%%%%%
%%%%%%%%%%%%%%%%%%
\begin{equation}
 N^T M^{(0)} N = N'^T U^T  M^{(0)}  U  N'  =  N'^T D N'
\label{k3}
\end{equation}
%%%%%%%%%%%%%%%%%%%%%%%%%%%%%%%%%%%%%%%%%%%%%
%%%%%%%%%%%%%%%%%%%
where $D=U^T M^{(0)} U$ and is given by
%%%%%%%%%%%%%%%%%%%%%%%%%%%%%%%%%%%%%%%%%%%%%
%%%%%%%%%%%%%%%%%%%
\begin{equation}
D = \left(
     \begin{array}{ccc}
      |A|&0&0  \\
      0&|A|&0   \\
      0&0&|B|\
     \end{array}
     \right)
\end{equation}
%%%%%%%%%%%%%%%%%%%%%%%%%%%%%%%%%%%%%%%%%%%%%
%%%%%%%%%%%%%%%%%%%

To determine whether or not there are any useful CP-violating phases
to
generate
$L \neq 0$ from (\ref{S8}), we consider the $\nu_L - N'$ Higgs
Yukawas from
(\ref{k2p})
%%%%%%%%%%%%%%%%%%%%%%%%%%%%%%%%%%%%%%%%%%%%%
%%%%%%%%%%%%%%%%%%
\begin{equation}
N^T M^{(1/2)}\nu_L  = N'^T U^T M^{(1/2)}\nu_L
\end{equation}
%%%%%%%%%%%%%%%%%%%%%%%%%%%%%%%%%%%%%%%%%%%%%
%%%%%%%%%%%%%%%%%%
and the Yukawa couplings $\lambda_{ij}$ can be simply read off
%%%%%%%%%%%%%%%%%%%%%%%%%%%%%%%%%%%%%%%%%%%%%
%%%%%%%%%%%%%%%%%%
\begin{equation}
\lambda_{ij} = {1 \over \sqrt{2}} \left( \begin{array}{ccc}
    3iue^{i\theta} & -3iue^{i\theta}  &  3ixe^{i\theta}   \\
   -3ue^{i\theta}  & -3ue^{i\theta}   &   -3xe^{i\theta}  \\
    0              &  -3\sqrt{2} x e^{i\phi} & -3\sqrt{2} w e^{i\phi}
\end{array}
\right)
\end{equation}
%%%%%%%%%%%%%%%%%%%%%%%%%%%%%%%%%%%%%%%%%%%%%
%%%%%%%%%%%%%%%%%%%
By direct computation we find,
%%%%%%%%%%%%%%%%%%%%%%%%%%%%%%%%%%%%%%%%%%%%%
%%%%%%%%%%%%%%%%%%%
\begin{eqnarray}
\epsilon_1 = \frac{1}{2\pi} \frac{9m(z)}{(2u^{\ast}u+x^{\ast}x)}
Im\left[(-iu^{\ast}x+ix^{\ast}w)^2 e^{2i(\phi-\theta)}\right]  \\
\epsilon_2 = \frac{1}{2\pi} \frac{9m(z)}{(2u^{\ast}u+x^{\ast}x)}
Im\left[(u^{\ast}x+x^{\ast}w)^2 e^{2i(\phi-\theta)}\right]~~~~  \\
\epsilon_3 = \frac{1}{2\pi} \frac{18m(1/z)}{(w^{\ast}w+x^{\ast}x)}
Im\left[x^{\ast}uw^{\ast}x e^{2i(\theta-\phi)}\right]~~~~~~~~~~~~
\end{eqnarray}
%%%%%%%%%%%%%%%%%%%%%%%%%%%%%%%%%%%%%%%%%%%%%
%%%%%%%%%%%%%%%%%%%
 where $z \equiv (M_3^2/M_1^2)$ and $m(z)$ is $f(z)$ or $g(z)$
(equations (\ref{ssf}) and (\ref{ssg}) respectively)
for the non-supersymmetric
or supersymmetric case.
As one can see there are non-vanishing phases for CP-violation.

We also look at the more restrictive ansatz of DHR, where only the
vev from a
single Higgs multiplet contributes to an element of the mass matrix.
In this
case $u$ and $w$ are proportional to $\bar{A}$ and $\bar{B}$
respectively, and
carry the same phases. So in this case we have:
%%%%%%%%%%%%%%%%%%%%%%%%%%%%%%%%%%%%%%%%%%%%%
%%%%%%%%%%%%%%%%%%%
\begin{eqnarray}
\epsilon_1 = \frac{1}{2\pi} \frac{9m(z)}{(2u^{\ast}u+x^{\ast}x)}
Im\left[-|u|^2x^2 e^{2i(\phi+\theta)}-|w|^2{x^{\ast}}^2
e^{2i(\phi+\theta)}\right]  \\
\epsilon_2 = - \epsilon_1~~~~~~~~~~~~~~~~~~~~~~~~~~~\\
\epsilon_3 = 0~~~~~~~~~~~~~~~~~~~~~~~~~~~~~
\end{eqnarray}
%%%%%%%%%%%%%%%%%%%%%%%%%%%%%%%%%%%%%%%%%%%%%
%%%%%%%%%%%%%%%%%%%%
Note that even in this special case the CP violating asymmetry is
non-vanishing, and depends on phases which are not fixed by low
energy
measurements.

In this case one also notes that the asymmetry is equal and opposite
for the
two, degenerate, lighter right-handed neutrino mass eigenstates. This
means
that if after inflation one reheated above the mass of these states,
populating
them thermally (hence equally, since they are degenerate), then their
resulting
decays would produce cancelling asymmetries, to the order in which we
have
calculated. The cancellation would presumably be vitiated by
differences in the
asymmetries generated by interferences at two loops. On
the other
hand, in the case where the two mass eigenstates are populated by
inflaton
decay, there is no reason to expect equal numbers of them to be
produced. Even
if the inflaton decays to N states in a way that is completely
flavour
symmetric in the chiral basis (unprimed), after the rotation to the
mass
eigenbasis there will be in general a difference in the numbers
produced
(unless in the initial chiral basis the flavour dependence in the
decay
mechanism was fine-tuned to be diagonal). In general, then, the
resulting
asymmetry will depend on the fraction of the inflaton decays to the
first N
mass eigenstate, minus the fraction of decays to the second,
degenerate, N mass
eigenstate. Let us denote this difference of fractions by $y$.

To get an idea of how large an asymmetry might be produced with this
mechanism
with these couplings, let us assume that the undetermined phases
appearing in
$\epsilon_1$ are generically of order unity. DHR have done fits to
the solar
neutrino data (and extracted the corresponding laboratory
predictions) with a
neutrino mass ansatze of this type (their ``case 1" for which they can
fit the
combined data). In particular the $\Delta{I_W}=(1/2)$ mass terms are
related to
the charge $2/3$ quark masses, and the parameters extracted from
their fits to
quark masses. The quark mass fits indicate $w>>x>>u$  meaning $z =
|w|^2/|u|^2>>1$
and with $m(z)\sim 1/ \sqrt{z}$ for large z, we find for maximal
phases that
%%%%%%%%%%%%%%%%%%%%%%%%%%%%%%%%%%%%%%%%%%%%%
%%%%%%%%%%%%%%%%%%
\begin{equation}
\epsilon_1 \sim  {\rm few} |u| |w|
\end{equation}
%%%%%%%%%%%%%%%%%%%%%%%%%%%%%%%%%%%%%%%%%%%%%
%%%%%%%%%%%%%%%%%%
the quark mass fits indicate: $|w| \simeq 1$, and $|u| \simeq .5
\times
10^{-3}$, giving $\epsilon_1 \sim 10^{-3}$ for maximal phases. The
MSW fit to
the solar neutrino data indicates a ${N'}_1$ mass of a few times
$10^{10} GeV$.
This is consistent with ${N'}_1$ production in inflaton decay, given
our
estimate of the inflaton mass from COBE of order $10^{11} GeV$. With
our
estimates for the inflaton mass and reheat temperature we then find:
%%%%%%%%%%%%%%%%%%%%%%%%%%%%%%%%%%%%%%%%%%%%%
%%%%%%%%%%%%%%%%%%
\begin{equation}
\frac{n_{B-\bar{B}}}{n_\gamma} \sim 10^{-4} y \epsilon
\end{equation}
%%%%%%%%%%%%%%%%%%%%%%%%%%%%%%%%%%%%%%%%%%%%%
%%%%%%%%%%%%%%%%%%
which for maximal phases gives sufficient  baryogenesis for $y$ as
small as
$10^{-4}$. So with parameters determined from  predictive ansatze for
neutrino
masses, and fit to the solar neutrino MSW oscillations, one can have
sufficient
baryogenesis by the Fukugita-Yanagida mechanism (see also
\cite{fy3}).
%%%%%%%%%%%%%%%%%%%%%%%%%%%%%%%%%%%%%%%%%%%%%
%%%%%%%%%%%%%%%%%%%%%

        Turning now to the scenario proposed in CDO for (s)neutrino
mass
induced
generation of a BAU, we find that there may be significant
differences in the efficacy of the mechanism, depending on how the
mass matrix texture has been implemented in the (unified) theory. The
issues involve the availability of the couplings required to induce the lepton
number violating scalar condensate potential, and finding flat directions along
which one can have
scalar condensate oscillations.

Since the efficiency of flat
direction oscillation mechanisms for baryogenesis in inflationary
cosmologies depends in part on the fact that inflationary
fluctuations will in general drive the scalars producing the operator
vev, to vevs $\phi_o \simeq \mu$ of the same order as the unification
scale, we must demand F-flatness including all superpotential
couplings to GUT sector fields. As one includes more Higgs
representations, and larger Higgs representations, which couple to
the matter 16s, one increases the number of F-terms whose vanishing
is required for the existence of the desired flat directions. As one
assembles generations of matter multiplets in single irreducible
representations of the gauge group, one reduces the number of
independent relative generation rotations one might perform to avoid
F-terms. So the more extensive the GUT sector structure, the more
difficult it becomes to find useful flat directions.

	We see this clearly in the class of models on which we are
focusing. If we were to consider the Yukawa coupling pattern with HRR
texture as just a pattern of couplings of the standard model
superfields, and restrict ourselves to only the standard model
multiplets, then we would have (among others) the flat direction
which we discussed in the previous section. Assuming the relation
between neutrino Dirac mass Yukawas and up quark mass Yukawas
implicit in this pattern (though now without any GUT sector), and
setting the scale of right handed neutrino masses of order $10^{11}$
GeV, as suggested by the MSW solution of the solar neutrino puzzle
\cite{dhr},and observable sector supersymmetry breaking scalar masses
of order 100 GeV, we simply plug into equation (\ref{g}) to determine the
resulting BAU:
%%%%%%%%%%%%%%%%%%%%%%%%%%%%%%%%%%%%%%%%%%%%%
%%%%%%%%%%%%
\begin{equation}
\frac{n_B}{n_\gamma} \simeq  O(10^{-2}) \frac{
\theta g_{\nu}^e g_{\nu}^\mu \mu^3 M^2}{{M_P}^4m_\delta}
\sim O(10^{-19})\theta
\label{gg}
\end{equation}
%%%%%%%%%%%%%%%%%%%%%%%%%%%%%%%%%%%%%%%%%
which is clearly insufficient. The origin of the numerical deficiency is the
identification $M \lappeq 10^{11}$GeV, which we made in order to
match our ansatze to the MSW solution of the solar neutrino problem.
Should we relax the requirement of MSW resolution of the neutrino puzzle then
we can easily raise $M$ sufficiently to provide baryogenesis within
our ansatze, with the $\nu_L$ see-sawing to unobservably light masses.

       Consider now the full SO(10)theory, as elaborated by HRR. Its
superpotential includes the terms  shown in equation (\ref{superp}),
responsible for the generation of standard model fermion masses, plus
an extensive set of Higgs self couplings (not shown) which are
required to generate the neccessary symmetry breaking vevs. The first thing
that we note is that since the ``right-handed'' neutrino now appears in a 16 of
SO(10), the ${\bf N^3}$ and ${\bf N H_1 H_2}$superpotential terms previously
used in generating the lepton number violating condensate potential terms, are
now forbidden by gauge invariance. However, with the introduction of an SO(10)
singlet coupling to itself and the Higgs 10, one has interactions that induce
(at four loops!) the required operator.  For flat directions, we do not
consider ones involving Higgs vevs, as they would be
required to satisfy F-flatness conditions associated with the GUT
Higgs self couplings which are very model dependent and which we have
not listed, and also as they risk vitiating the generation trick on
which
we will be forced to rely on below, to avoid F-terms when giving vevs
to chiral scalars in the 16s. That we will be forced to rely on a
generation trick when introducing vevs in the 16s can be seen as
follows. In the superpotential Yukawas there are couplings
$16\cdot16\cdot\bar{126}$ for each of the  $\bar{126}$s. Let us
consider the coupling of a $\bar{126}$ to a single (generation of)
16. Now consider a particular component of the 16 to which we wish to
give a vev. By a choice of basis for the subgroup decomposition we
may choose this state in the 16 to be the singlet in the
SO(10)$\supset$ SU(5)xU(1) decomposition 16 = 1+$\bar{5}$+10. Under
SU(5)xU(1) decomposition the $\bar{126}$ also contains a singlet.
Furthermore, by inspection of the decomposition of the tensor product
$16\cdot16\cdot\bar{126}$ under SU(5)xU(1) it is clear that the
singlet in the $\bar{126}$ is that appearing (with the same choice of
subalgebra basis) in the product of singlets in the 16s. Since by
hypothesis this basis was chosen such that the singlet of the 16 was
the component receiving a vev, this means that the F-term for the
singlet component of the $\bar{126}$ is nonvanishing, as in the
product it is only coupled to the nonvanishing singlets in the 16. In
order to evade this argument, we require several generations of 16,
such that some linear combination of them has vanishing (generation)
diagonal coupling to itself. Inspecting the pattern of superpotential
Yukawas in the HRR GUT [equation (\ref{superp})] we observe that this is
exactly what occurs for the ${16}_{1}$; the discrete symmetry used to
enforce the HRR texture has been precisely arrange to prevent its
self coupling. So provided vevs in the other 16s, and the Higgs
representations, vanish, any pattern of vevs in the ${16}_{1}$ will
be F-flat. It only remains to satisfy the D-flatness condition. As
shown in \cite{ad}, we only have to consider D-terms associated with
generators unbroken at the GUT scale. Since the standard model gauge
group has 12 generators, there are 12 non-trivial quadratic equations that
must be satisfied for D-flatness. We have 16 complex scalars in our
generation; the SU(5)xU(1) singlet is singlet under the standard
model generators (in the basis where the standard model group is
embedded in the SU(5)) and hence the vev for the singlet receives no
D-term contributions. For the remaining 15 chiral scalars we have 12
conditions, hence we expect the solution space to be parametrized by
the vevs of any three of the scalars, with the other vevs being
continuous, in general non-vanishing, functions of these three.

     If one is looking for a vev for a B-L$\neq$0 gauge invariant
quartic scalar operator, as an induced potential term that will
generate nonzero B-L so the resulting BAU is not erased by sphaleron
effects, then those formed from standard model supermultiplets have
been listed by Morgan \cite{morg}. Those with nonzero B-L all involve
Higgs vevs, except for the operator we used in our example in the
previous section, so we must use it again in this case. (We should be
careful here as there may be new operators that can be induced since
a singlet right handed neutrino has a vev; we leave consideration of
these  as
an
exercise for the reader). In the case of our single generation flat
direction (pure ${16}_1$) the vev arises when all the external fields
lie in the ${16}_1$. In this basis the Yukawas are not generation
diagonal, and the net effect is that in the diagram inducing the
operator with all  ${16}_1$ external legs, instead of factors of the
top and bottom Yukawas, we get two factors of Yukawas whose magnitude
is essentially the geometric mean of the first and second generation
Yukawas; ie we have traded the coupling factors ``down" in generation.
In our induced operator, and hence in the final BAU produced, this
results in a suppression of four or five orders of magnitude,
which when compounded with the fact that our induced operator now appears at
four loop order implies that even with $N$ masses raised toward the unification
scale, this
mechanism of baryogenesis would be, in this ansatze, of  questionable
viability.

     While we have concentrated in this section on SO(10) inspired
ansatze of the HRR type, similar issues will arise in other models
incorporating a neutrino mass see-saw, such as the SU(5)xU(1) models
\cite{eno5}. One should also bear in mind that the absence of B-L
violation in gauge induced operators in many supersymmetric GUTs,
means superpotential interactions such as neutrino masses may be the
only way to produce a B-L from flat direction oscillations in many
supersymmetric GUT models, and hence the only way to produce a BAU
not subject to sphaleron erasure. As such, this possibility merits
careful consideration in all prospective unified models.
%%%%%%%%%%%%%%%%%%%%%%
\section{Conclusions}

%%%%%%%%%%%%%%%%%%%%%%%%%%%%%%%%%%%%%%%%%%%%%
%%%%%%%%%%%%%%%%%%%%%%%%
       The recent COBE observations, as well as giving us information
about the
inflationary epoch, define for us the cosmological parameters within
which
post-inflationary scenarios of baryogenesis must operate. For general
classes
of models, if one eschews fine tuning of parameters, we showed that
the COBE
observations indicate an inflaton mass of order $10^{11}GeV$ and
suggests a
reheat temperature of order $10^8 GeV$. These low values are
constraining on
models of baryogenesis; in particular conventional GUT
out-of-equilibrium decay
scenarios have difficulty reconciling this with the observed
stability of the
proton. In view of this, we examined alternative mechanisms for
baryogenesis,
where a lepton asymmetry is created before the electroweak phase
transition,
and then partially reprocessed into baryons by nonperturbative
electroweak
effects. In both of these scenarios the lepton violating interaction
generating
the lepton asymmetry is the Majorana mass of a right-handed see-saw
(s)neutrino. In the first mechanism, proposed by Fukugita and
Yanagida, the
lepton asymmetry is generated by the out of equilibrium decay of the
right-handed neutrino. We have calculated the generation of lepton
and baryon
asymmetries via this mechanism in supersymmetric models, with
sneutrino as well
as neutrino decay, and shown that it proceeds unsuppressed, as in the
non-supersymmetric case. We have also discussed another mechanism
which we
recently proposed \cite{cdo1}for supersymmetric models with see-saw
neutrino
masses, in which the Majorana masses induce effective
lepton-number-violating
potential terms, which act on condensate oscillations of squarks and
sleptons
along ``flat" directions of the scalar potential. Although this
mechanism only
operates in supersymmetric models, it does have the advantage of
being
operative even in the case where the right handed (s)neutrinos are
too massive
to be produced after inflation. Finally, we examined the
implementation of
these mechanisms in the context of recently proposed ansatze for
neutrino
masses. The Fukugita-Yanagida mechanism was, in principle, amply able to
generate a
sufficient
baryon asymmetry, though the actual result could not be
definitely predicted; this was because phases entered which could not be
measured at low
energies. On the other hand,with this particular ansatze for neutrino masses,
 the mechanism proposed in CDO would not produce a sufficient BAU if one
simultaneously demanded that one resolve the solar neutrino problem by MSW
oscillation; this was because the MSW fit in this ansatze gave too low a
Majorana mass to the right-handed neutrino. In models (and fits) with
right-handed
neutrino masses closer to the unification scale, this mechanism would be
more efficient, and could function well (although not in the SO(10) models we
have considered) when right-handed $N$s are too
heavy to be directly produced in inflaton decay, as required by the
Fukugita-Yanagida mechanism.
 In conclusion, these may be  efficient mechanisms for
baryogenesis,
which we naturally expect to arise in models incorporating realistic
see-saw
masses for neutrinos, and represent a viable  option for
cosmic
baryogenesis.

%%%%%%%%%%%%%%%%%%%%%%%%%%%%%%%%%%%%%%%%%%%%%
%%%%%%%%%%%%%%%%%%%%%%%%

\vskip .1in

\noindent{ {\bf Acknowledgements} } \\
\noindent We thank M.-K. Gaillard, Marcus Luty, Lawrence Hall and Mark
Srednicki
for very
helpful discussions. The
work of BAC and SD was supported in part by the Natural Sciences and
Engineering Research council of Canada. The work of KAO was supported
in part by DOE grant DE-AC02-83ER-40105, and by a Presidential Young
Investigator Award. BAC and KAO would like to thank the CERN Theory
Division for kind hospitality during part of this research.

\newpage

\begin{center}
{\large{\bf Appendix}}
\end{center}
In this appendix, we wish to discuss the calculation of the
CP-violating
parameter $\epsilon$ in  more detail. Since flipping a small number
of signs
would make $\epsilon$ zero, one must be sure that the loop diagrams
of figure
2 do not correspond to an F-term vertex, because
non-renormalisation theorems say that only D-terms can have loop
corrections.
To distinguish scalars from fermions in this section, the scalar
component
fields will have tildes: {\bf N} = $\tilde{N} + \theta N +\theta
\theta
F_{N}$.

In a flavour basis for the ${\bf L}^i$, there will clearly be no
gauge particle
contributions to CP violation at one loop, so one only needs to worry
about
diagrams involving the components of the ${\bf N}^j, {\bf H}$ and
${\bf L}^i$
superfields.
We neglect supersymmetry-breaking and Higgs mixing masses, on the
assumption that these are small compared to $M_i$, so ${\bf L}^i$ and
${\bf H}$
 only have D-term propagators: if there are no mass terms $m_i {\bf
L}^i {\bf
L}^i$,
 $ \mu {\bf H}{\bf H}$ in the superpotential, then only the $<{\bf
L}{\bf
L}^{\dagger}>$
 and $< {\bf H}{\bf H}^{\dagger}>$ two-point functions will appear.
This of
course is
not the case for the ${\bf N}$ superfield, which has both F-term
($<{\bf N}{\bf
N}>$)
and D-term ($<{\bf N}{\bf N}^{\dagger}>$) propagators. However, since
the
component
diagrams of figure 2  have mass insertions on the $\tilde{N}$
and $N$ lines, one would expect $<{\bf N}^i{\bf N}^i> \sim M_i^*
\Delta_F$ to
appear.
This is in fact the case, as one can see from figure 4, the
supergraph corresponding to the tree $\times$ loop$^{\dagger}$
matrix elements used to compute $\epsilon$. As previously claimed,
 the loop is a correction to the D-term ${\bf N} {\bf
H}^{\dagger}{\bf
L}^{\dagger}$,
 so is not subject to the non-renormalisation theorems.

To calculate the lepton asymmetry produced in the decay of the
$\tilde{N^i}$
 and $N^i$, one only needs the CP violating difference $\epsilon$,
which is
proportional to the imaginary part of the coupling constant
combination
times the imaginary part of the matrix element squared. The only
kinematically
allowed way to have an on-shell intermediate state particle
(neccesary for the
[matrix element]$^2$ to develop an imaginary part) is to put the
${\bf H}$ and
the ${\bf L}$ in the loop on-shell, and the ${\bf N}$ off-shell. It
is then
easy to compute
$\epsilon$ for each decay. The combination of coupling constants that
needs
to have a phase, for a lepton asymmetry to be produced in the decay of
$\tilde{N}^i$ and $N^i$, can be read from the supergraph to be
%%%%%%%%%%%%%%%%%%%%%%%%%%%%%%%%%%%%%%%%%%%%%
%%%%%%%%%%%%%%%%%%%%%
\begin{eqnarray*}
{}~~~~~~~~~~~~~~~~~~~~~\sum_j M_i^* (\lambda \lambda^{\dagger})_{ij}
M_j
(\lambda ^*
\lambda^{T})_{ji}~~~~~~~~~~~~~~~~~~~~~~~~~~~~~~(A1)
\end{eqnarray*}
%%%%%%%%%%%%%%%%%%%%%%%%%%%%%%%%%%%%%%%%%%%%%
%%%%%%%%%%%%%%%%%%%%%
At first sight the complex-conjugates on the masses appear backwards,
and
not terribly important in any case because one can always work in a
basis
where the masses are real. However, since propagators appear in
amplitudes, and inverse propagators in the Lagrangian. $M_i^* M_j$ is
correct. One can also check that (A1) is invariant under phase
rotations
of the ${\bf N}^i$, which is good, because a physical quantity
($\epsilon$)
should
not depend on phase choices. It is also of some interest to get the
form of
(A1) in an arbitrary basis for the ${\bf N}^i$, because most models
do not
 predict a real diagonal majorana mass matrix, and the algebra
neccessary
to check if there is a phase is simplified if one does not have to
rotate
 to the basis where $M_i \in {\cal R}$. If $\hat{M}$ is a symmetric
majorana
mass matrix, and $\hat{\lambda}$ is the yukawa coupling in this
basis, then
$\hat{M}$ can be diagonalized in this basis by a unitary matrix $U$:
%%%%%%%%%%%%%%%%%%%%%%%%%%%%%%%%%%%%%%%%%%%%
\begin{eqnarray*}
{}~~~~~~~~~~~~~~~~~~~~U^{T} \hat{M} U = diag[M_1, M_2,
M_3]~~~~~~~~~~~~~~~~~~~~~~~~~~~~~(A2)
\end{eqnarray*}
%%%%%%%%%%%%%%%%%%%%%%%%%%%%%%%%%%%%%%%%%%%%%
The coupling constant combination that needs a phase is therefore
%%%%%%%%%%%%%%%%%%%%%%%%%%%%%%%%%%%%%%%%%%%%%
%%%
\begin{eqnarray*}
{}~~~~~~~~~~~~~~~~~~~~~~~\left( U^{\dagger}
\hat{M}^{\dagger}\hat{\lambda}\hat{\lambda}^{\dagger}\hat{M}
\hat{\lambda}^*\hat{\lambda}^T U \right)_{ii} \label{S11}
{}~~~~~~~~~~~~~~~~~~~~~~~~~~~~~~~~~~(A3)
\end{eqnarray*}
%%%%%%%%%%%%%%%%%%%%%%%%%%%%%%%%%%%%%%%%%%%%%
%%
Note that this is true in any basis, provided that $U$
diagonalizes the majorana mass matrix.

%%%%%%%%%%

%%%%%%%%%%%%%%%%%%%%%%%%%%%%%%%%%%%%%%%%%%%%%
%%%%%%%%%%%%%%%%%%%%%%%%

%%%%%%%%%%
%%%%%%%%%%%%%%%%%%%%%%%%%%%%%%%%%%%%%%%%%%%%%
%%%%%%%%%%%%%%%%%%%%%%%%

%%%%%%%%%%

}
}

\newpage

\input FEYNMAN
\bigphotons
%dimension five thing
\begin{picture}(40000,20000)
\put(8000,0){$\tilde{Q}$}
\drawline\scalar[\SE\REG](10000,0)[4]
\drawline\fermion[\SW\REG](\pbackx,\pbacky)[8000]
\global\advance\pbackx by -2000
\put(\pbackx,\pbacky){$Q$}
\drawline\fermion[\E\REG](\pfrontx,\pfronty)[10000]
\global\advance\pmidy by -250
\put(\pmidx,\pmidy){$\times$}
\global\advance\pmidx by -2500
\global\advance\pmidy by 600
\put(\pmidx,\pmidy){$\tilde{H_1}$}
\global\advance\pmidx by 5000
\put(\pmidx,\pmidy){$\tilde{H_2}$}
\drawline\fermion[\SE\REG](\pbackx,\pbacky)[8000]
\global\advance\pbackx by 1000
\put(\pbackx,\pbacky){$L$}
\drawline\scalar[\NE\REG](\pfrontx,\pfronty)[4]
\global\advance\pbackx by 1000
\put(\pbackx,\pbacky){$\tilde{Q}$}
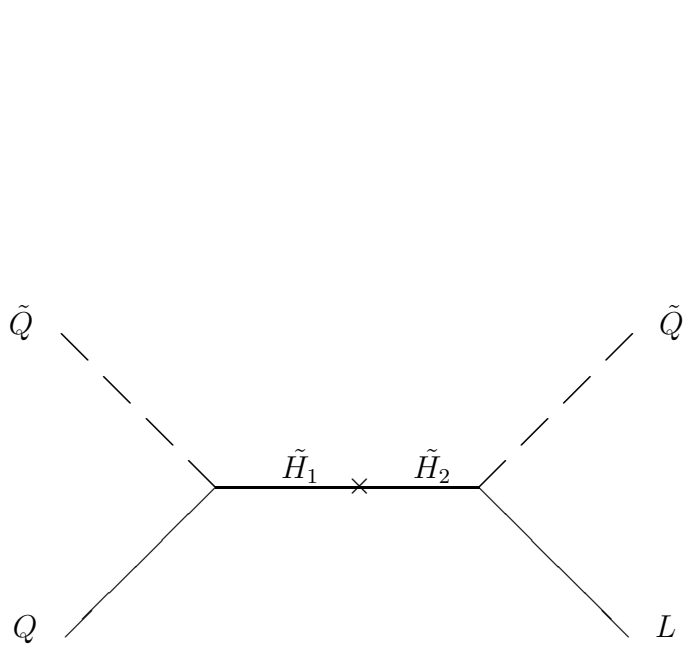
\put(0,-20000){ Figure 1:  Diagram inducing baryon number violating
dimension five operator in susy SU(5)}
\end{picture}

%four loops for susy F+Y
\begin{picture}(20000,50000)
\put(0,52000){a)}
\drawline\fermion[\E\REG](2000,52000)[8000]
\global\advance\pmidy by 650
\put(\pmidx,\pmidy){$N^i$}
\drawline\scalar[\SE\REG](\pbackx,\pbacky)[4]
\global\advance\pmidy by -1250
\global\advance\pmidx by -2400
\put(\pmidx,\pmidy){$\tilde L,H$}
\drawline\fermion[\NE\REG](\pfrontx,\pfronty)[8000]
\global\advance\pmidy by 700
\global\advance\pmidx by -1900
\put(\pmidx,\pmidy){$\tilde H,L$}
\drawline\scalar[\E\REG](\pbackx,\pbacky)[4]
\global\advance\pmidy by 500
\put(\pmidx,\pmidy){$\tilde L,H$}
\drawline\fermion[\S\REG](\pfrontx,\pfronty)[11330]
\global\advance\pmidx by -250
\put(\pmidx,\pmidy){$\times$}
\global\advance\pmidx by 1200
\put(\pmidx,\pmidy){$N^j$}
\drawline\fermion[\E\REG](\fermionbackx,\fermionbacky)[8000]
\global\advance\pmidy by 500
\put(\pmidx,\pmidy){$\tilde H,L$}

\put(0,37000){b)}
\drawline\fermion[\E\REG](2000,37000)[8000]
\global\advance\pmidy by 500
\put(\pmidx,\pmidy){$N^i$}
\drawline\fermion[\SE\REG](\pbackx,\pbacky)[8000]
\global\advance\pmidy by -1250
\global\advance\pmidx by -2400
\put(\pmidx,\pmidy){$L,\tilde H$}
\drawline\scalar[\NE\REG](\pfrontx,\pfronty)[4]
\global\advance\pmidy by 700
\global\advance\pmidx by -1900
\put(\pmidx,\pmidy){$H,\tilde L$}
\global\advance\pbackx by -250
\global\advance\pbacky by -250
\put(\pbackx,\pbacky){$\times$}
\global\advance\pbackx by 250
\global\advance\pbacky by 250
\drawline\scalar[\E\REG](\pbackx,\pbacky)[4]
\global\advance\pmidy by 500
\put(\pmidx,\pmidy){$\tilde L,H$}
\drawline\scalar[\S\REG](\pfrontx,\pfronty)[5]
\global\advance\pmidx by 600
\put(\pmidx,\pmidy){$\tilde{N}^j$}
\drawline\fermion[\E\REG](\fermionbackx,\fermionbacky)[8000]
\global\advance\pmidy by 500
\put(\pmidx,\pmidy){$\tilde H,L$}

\put(0,22000){c)}
\drawline\scalar[\E\REG](2000,22000)[4]
\global\advance\pmidy by 500
\put(\pmidx,\pmidy){$\tilde{N}^i$}
\global\advance\pbackx by -250
\global\advance\pbacky by -250
\put(\pbackx,\pbacky){$\times$}
\global\advance\pbackx by 250
\global\advance\pbacky by 250
\drawline\scalar[\SE\REG](\pbackx,\pbacky)[4]
\global\advance\pmidy by -1250
\global\advance\pmidx by -2400
\put(\pmidx,\pmidy){$\tilde L,H$}
\drawline\scalar[\NE\REG](\pfrontx,\pfronty)[4]
\global\advance\pmidy by 700
\global\advance\pmidx by -1900
\put(\pmidx,\pmidy){$H,\tilde L$}
\drawline\fermion[\E\REG](\pbackx,\pbacky)[8000]
\global\advance\pmidy by 550
\put(\pmidx,\pmidy){$L,\tilde H$}
\drawline\fermion[\S\REG](\pfrontx,\pfronty)[11330]
\global\advance\pmidx by -200
\put(\pmidx,\pmidy){$\times$}
\global\advance\pmidx by 1500
\put(\pmidx,\pmidy){$N^j$}
\drawline\fermion[\E\REG](\pbackx,\pbacky)[8000]
\global\advance\pmidy by 500
\put(\pmidx,\pmidy){$\tilde H,L$}

\put(0,7000){d)}
\drawline\scalar[\E\REG](2000,7000)[4]
\global\advance\pmidy by 500
\put(\pmidx,\pmidy){$\tilde{N}^i$}
\drawline\fermion[\SE\REG](\pbackx,\pbacky)[8000]
\global\advance\pmidy by -1250
\global\advance\pmidx by -2400
\put(\pmidx,\pmidy){$\tilde H,L$}
\drawline\fermion[\NE\REG](\pfrontx,\pfronty)[8000]
\global\advance\pmidy by 700
\global\advance\pmidx by -1900
\put(\pmidx,\pmidy){$L,\tilde H$}
\drawline\scalar[\E\REG](\pbackx,\pbacky)[4]
\global\advance\pmidy by 500
\put(\pmidx,\pmidy){$H,\tilde L$}
\drawline\fermion[\S\REG](\pfrontx,\pfronty)[11330]
\global\advance\pmidx by -200
\put(\pmidx,\pmidy){$\times$}
\global\advance\pmidx by 1500
\put(\pmidx,\pmidy){$N^j$}
\drawline\scalar[\E\REG](\pbackx,\pbacky)[4]
\global\advance\pmidy by 500
\put(\pmidx,\pmidy){$\tilde L,H$}
\global\advance\pmidy by -3500
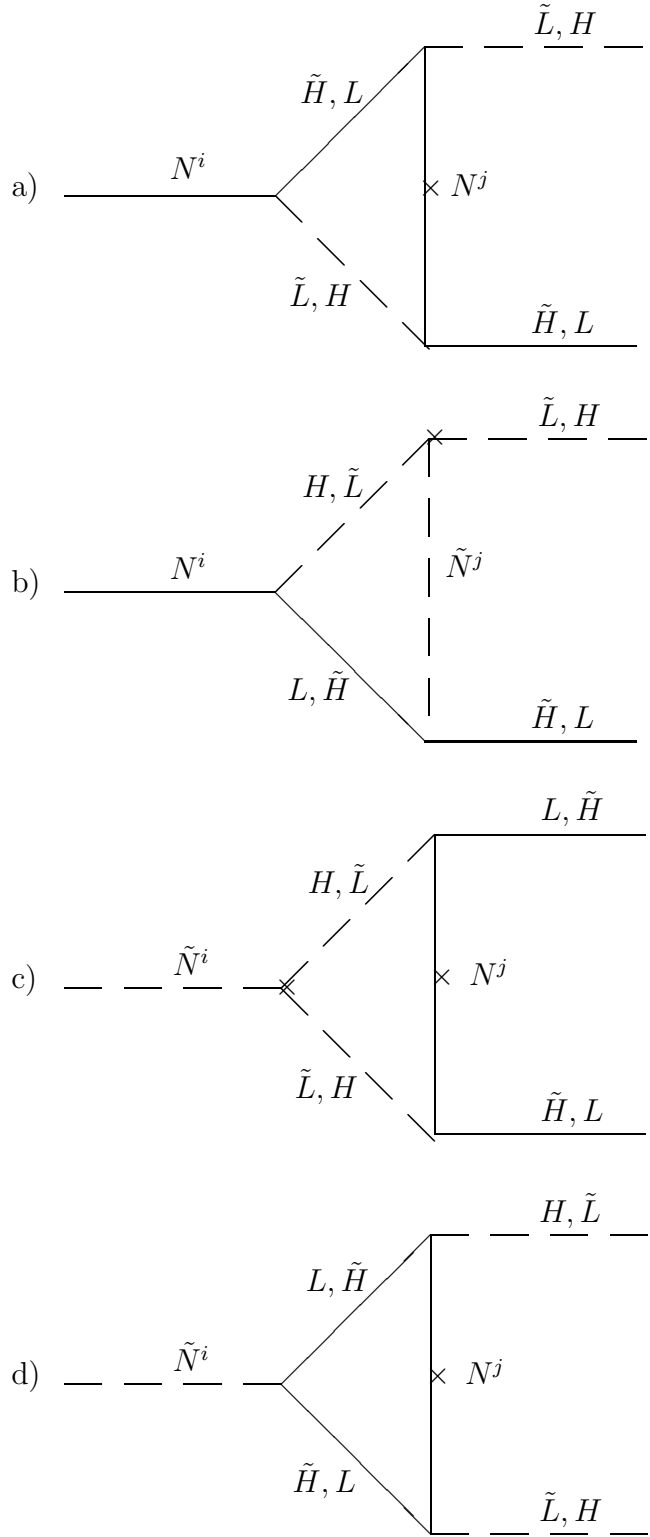
\put(0,\pmidy){Figure 2: One-loop diagrams contributing to CP violation in
neutrino and
sneutrino decays;}
\global\advance\pmidy by -1500
\put(0,\pmidy){only diagram a) with $H$ and $L$ will contribute in
the non-supersymmetric case}
\end{picture}

\begin{picture}(15000,10000)
\put(0,-35000){Figure 3:  Diagram inducing lepton number violating
potential terms from the neutrino mass}
\put(0,-36500){ see-saw superpotential after
supersymmetry breaking.}
\end{picture}

\newpage

%supergraph
\begin{picture}(40000,20000)
\drawline\fermion[\E\REG](0,0)[9000]
\global\advance\pbackx by -1500
\global\advance\pbacky by 500
\put(\pbackx,\pbacky){\bf N$^i$}
\global\advance\pbackx by 3000
\global\advance\pbacky by 2500
\put(\pbackx,\pbacky){\bf L$^k$}
\global\advance\pbacky by -6500
\put(\pbackx,\pbacky){\bf H}
\global\advance\pbacky by 3000
\put(\pbackx,\pbacky){$\lambda_{ik}$}
\global\advance\pbackx by -1500
\global\advance\pbacky by 500
\drawline\fermion[\SE\REG](\pbackx,\pbacky)[18000]
\global\advance\pbackx by -4000
\global\advance\pbacky by -1200
\put(\pbackx,\pbacky){\bf H$^{\dagger}$}
\global\advance\pbackx by 3300
\global\advance\pbacky by -500
\put(\pbackx,\pbacky){$\lambda^*_{jl}$}
\global\advance\pbacky by 500
\global\advance\pbackx by 3500
\put(\pbackx,\pbacky){\bf L$^{l \dagger}$}
\global\advance\pbacky by 4200
\global\advance\pbackx by -1800
\put(\pbackx,\pbacky){\bf N$^{j \dagger}$}
\global\advance\pbackx by -1000
\global\advance\pbacky by -3000
\drawline\fermion[\N\REG](\pbackx,\pbacky)[25500]
\global\advance\pmidx by -250
\put(\pmidx,\pmidy){$\times$}
\global\advance\pmidx by 1500
\put(\pmidx,\pmidy){$M_j$}
\global\advance\pmidx by -1500
\global\advance\pbackx by -4000
\put(\pbackx,\pbacky){\bf L$^{k \dagger}$}
\global\advance\pbackx by 3300
\global\advance\pbacky by 500
\put(\pbackx,\pbacky){$\lambda^*_{jk}$}
\global\advance\pbacky by -500
\global\advance\pbackx by 2500
\put(\pbackx,\pbacky){\bf H$^{\dagger}$}
\global\advance\pbacky by -3500
\global\advance\pbackx by -1000
\put(\pbackx,\pbacky){\bf N$^{j \dagger}$}
\global\advance\pbackx by -800
\global\advance\pbacky by 3500
\drawline\fermion[\SW\REG](\pbackx,\pbacky)[18000]
\drawline\fermion[\SE\REG](\pfrontx,\pfronty)[18000]
\global\advance\pbackx by 1000
\global\advance\pbacky by 500
\put(\pbackx,\pbacky){\bf N$^i$}
\global\advance\pbackx by -4000
\global\advance\pbacky by -1000
\put(\pbackx,\pbacky){$\lambda_{il}$}
\global\advance\pbackx by 1000
\global\advance\pbacky by 4500
\put(\pbackx,\pbacky){\bf H}
\global\advance\pbacky by -7500
\put(\pbackx,\pbacky){\bf L$^l$}
\global\advance\pbackx by 2000
\global\advance\pbacky by 3500
\drawline\fermion[\E\REG](\pbackx,\pbacky)[15000]
\global\advance\pmidy by -250
\put(\pmidx,\pmidy){$\times$}
\global\advance\pmidy by 1200
\put(\pmidx,\pmidy){$M_i^*$}
\drawline\fermion[\SW\REG](\pfrontx,\pfronty)[18000]
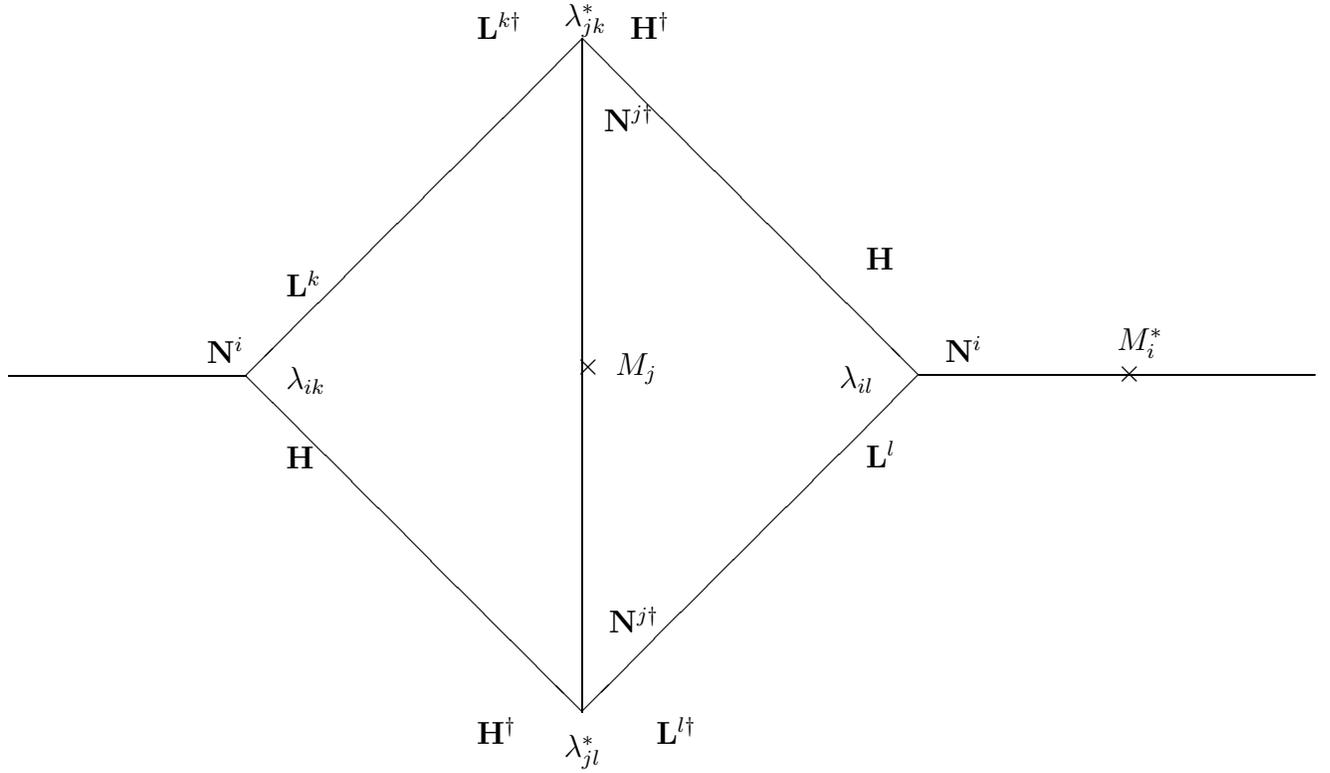
\put(0,-18000){Figure 4: supergraph corresponding to the loop $\times$
tree$^{\dagger}$ matrix
elements providing  }
\put(0,-19500){CP violation in heavy singlet neutrino and sneutrino decays}
\end{picture}
\end{document}